\documentclass[twocolumn,preprintnumbers,amsmath,amssymb]{revtex4}
\usepackage{graphicx}
\usepackage{dcolumn}
\usepackage{bm}

\begin{document}

\title{Simulation of traffic flow at a signalised intersection}

\author{M. Ebrahim Foulaadvand ~\footnote{Corresponding author: foolad@iasbs.ac.ir}}
\affiliation{Department of Nano-Science,
 Institute for Studies in Theoretical Physics and Mathematics (IPM),
P.O. Box 19395- 5531,Tehran, Iran and Department of Physics, Zanjan
University, P.O. Box 45195-313, Zanjan,  Iran.}
\author{Sommayeh Belbaasi}
\affiliation{Department of Physics, Zanjan University, P.O. Box
45195-313, Zanjan,  Iran.}

\date{\today}
\begin{abstract}

We have developed a Nagel-Schreckenberg cellular automata model
for describing of vehicular traffic flow at a single
intersection. A set of traffic lights operating either in
fixed-time or traffic adaptive scheme controls the traffic flow.
Closed boundary condition is applied to the streets each of which
conduct a uni-directional flow. Extensive Monte Carlo simulations
are carried out to find the model characteristics. In particular,
we investigate the dependence of the flows on the signalisation
parameters.

\end{abstract}


\maketitle
\section{Introduction}

Modelling the dynamics of vehicular traffic flow by cellular
automata has constituted the subject of intensive research by
statistical physics during the past years
\cite{schadrev,helbingrev,kernerbook,tgf05}. {\it City traffic}
was an early simulation target for statistical physicists
\cite{bml,nagatani1,nagatani2,cuesta,freund,chopard,tadaki,torok,chau,cs,brockfeld}.
Evidently the optimisation of traffic flow at a single
intersection is a preliminary but crucial step to achieve the
ultimate task of global optimisation in city networks
\cite{chitur}. Nagatani proposed the first model for simulation
of two crossing streets \cite{nagatani3}. Subsequently, Ishibashi
and Fukui developed similar models \cite{ishibashi1,ishibashi2}.
Recently, physicists have notably attempted to simulate the
traffic flow at intersections and other traffic designations such
as roundabouts
\cite{foolad1,krbalek,foolad2,foolad3,foolad4,helbing1,helbing2,helbing3,ray,chen,wang,huang,najem,foolad5}.
In principle, the vehicular flow at the intersection of two roads
can be controlled via two distinctive schemes. In the first
scheme, the traffic is controlled without traffic lights
\cite{foolad5,foolad6}. In the second scheme, signalised traffic
lights control the flow. In Ref.\cite{foolad2}, we have modelled
the traffic flow at a single intersection with open boundary
conditions on streets. Dependence of waiting times on
signalisation and inflow rates were investigated in details.
Along this line of study, our objective in this paper is to study
in some depth, the characteristics of traffic flow and its
optimisation in a single intersection with closed boundary
condition. We will compare our results to those obtained in our
recent study of a nonsignalised intersection \cite{foolad5}. In
order to capture the basic features of this problem, we have
constructed a NS cellular automata model. This paper has the
following layout. In section II, the model is introduced and
formulated. In sections III and IV, the results of Monte Carlo
simulations of each controlling scheme are exhibited. Concluding
remarks and discussions ends the paper in section V.

\section{ Description of the Problem }

Imagine two perpendicular one dimensional closed chains each
having $L$ sites. The chains represent urban roads accommodating
unidirectional vehicular traffic flows. They intersect each other
at the middle sites $i_1=i_2=\frac{L}{2}$ on the first and the
second chain. With no loss of generality we take the flow
direction in the first chain from south to north and in the
second chain from east to west. (see Fig.1 for illustration).
Cars are not allowed to turn. The discretisation of space is such
that each car occupies an integer number of cells denoted by
$L_{car}$. The car position is denoted by the location of its
head cell. Time elapses in discrete steps of $\Delta t$ and
velocities take discrete values $0,1,2,\cdots, v_{max}$ in which
$v_{max}$ is the maximum velocity.

\begin{figure}
\centering
\includegraphics[width=7.5cm]{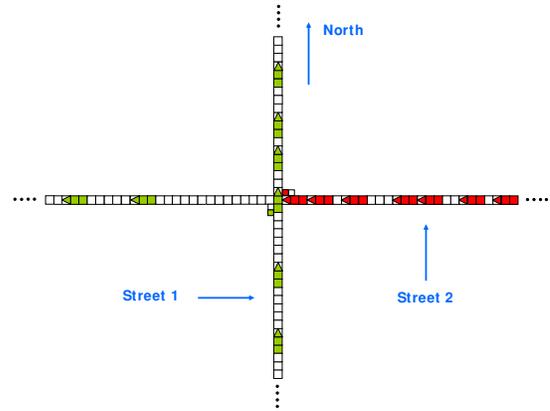}
\caption{ Intersection of two uni-directional streets. } \label{fig:bz2}
\end{figure}

To be more specific, at each step of time, the system is
characterized by the position and velocity configurations of
cars. The system evolves under the Nagel-Schreckenberg (NS)
dynamics \cite{ns}. Let us now specify the physical values of our
time and space units. The length of each car is taken 4.5 metres.
Therefore, the spatial grid $\Delta x$ (cell length) equals to
$\frac{4.5}{L_{car}}~m$. We take the time step $\Delta t=1~s$.
Furthermore, we adopt a speed-limit of $75~ km/h$. The value of
$v_{max}$ is so chosen to give the speed-limit value $75~km/h$ or
equivalently $21~m/s$ . In this regard, $v_{max}$ is given by the
integer part of $21 \times L_{car}/4.5$. For instance, for
$L_{car}=5$, $v_{max}$ equals $23$. In addition, each discrete
increments of velocity signifies a value of $\Delta
v=\frac{4.5}{L_{car}} m/s $ which is also equivalent to the
acceleration. For $L_{car}=5$ this gives the comfort acceleration
$0.9 ~\frac{m}{s^2}$. Moreover, we take the value of random
breaking parameter at $p=0.1$. A set of traffic lights controls
the traffic flow. We discuss two types of signalisation in this
paper: {\it fixed-time} and {\it traffic responsive}.

\section{Fixed Time Signalisation of lights}


In this scheme the lights periodically turn into red and green.
The period $T$, hereafter referred to as {\it cycle time}, is
divided into two phases. In the first phase with duration $T_g$,
the lights are green for the northward street and red for the
westward one. In the second phase which lasts for $T-T_g$
timsteps the lights change their colour i.e.; they become red for
the northward and green for the westward street. The gap of all
cars are update with their leader vehicle except those two which
are the nearest approaching cars to the intersection. These two
cars need special attention. For these approaching cars gap
should be adjusted with the signal in its red phase. In this case,
the gap is defined as the number of cells right after the car's
head to the intersection point $\frac{L}{2}$. If the head
position of the approaching car lies in the crossing point the
gap is zero. We note that at the time step when the traffic light
goes red for a direction, portion of a car from that direction
can occupy the crossing point. In this case the leading car of
the queue in the other direction should wait until the passing
car from the crossing point completely passes the intersection
i.e.; its tail cell position become larger than $\frac{L}{2}$.
Now all the computational ingredients for simulation is at our
disposal. The streets sizes are $L_1=L_2=1350~m$ and we take
$L_{car}=5$. The system is update for $2\times10^5$ time steps.
After transients, two streets maintain steady-state currents
denoted by $J_1$ and $J_2$ which are defined as the number of
vehicles passing from a fixed location per time step. They are
functions of global densities $\rho_1$ and $\rho_2$ and signal
times $T$ and $T_g$. We kept $\rho_2$ fixed in the second street
and varied $\rho_1$. Figure (2) exhibits the fundamental diagram
of the first street i.e.; $J_1$ versus $\rho_1$. The parameters
are specified in the caption.

\begin{figure}
\centering
\includegraphics[width=7cm]{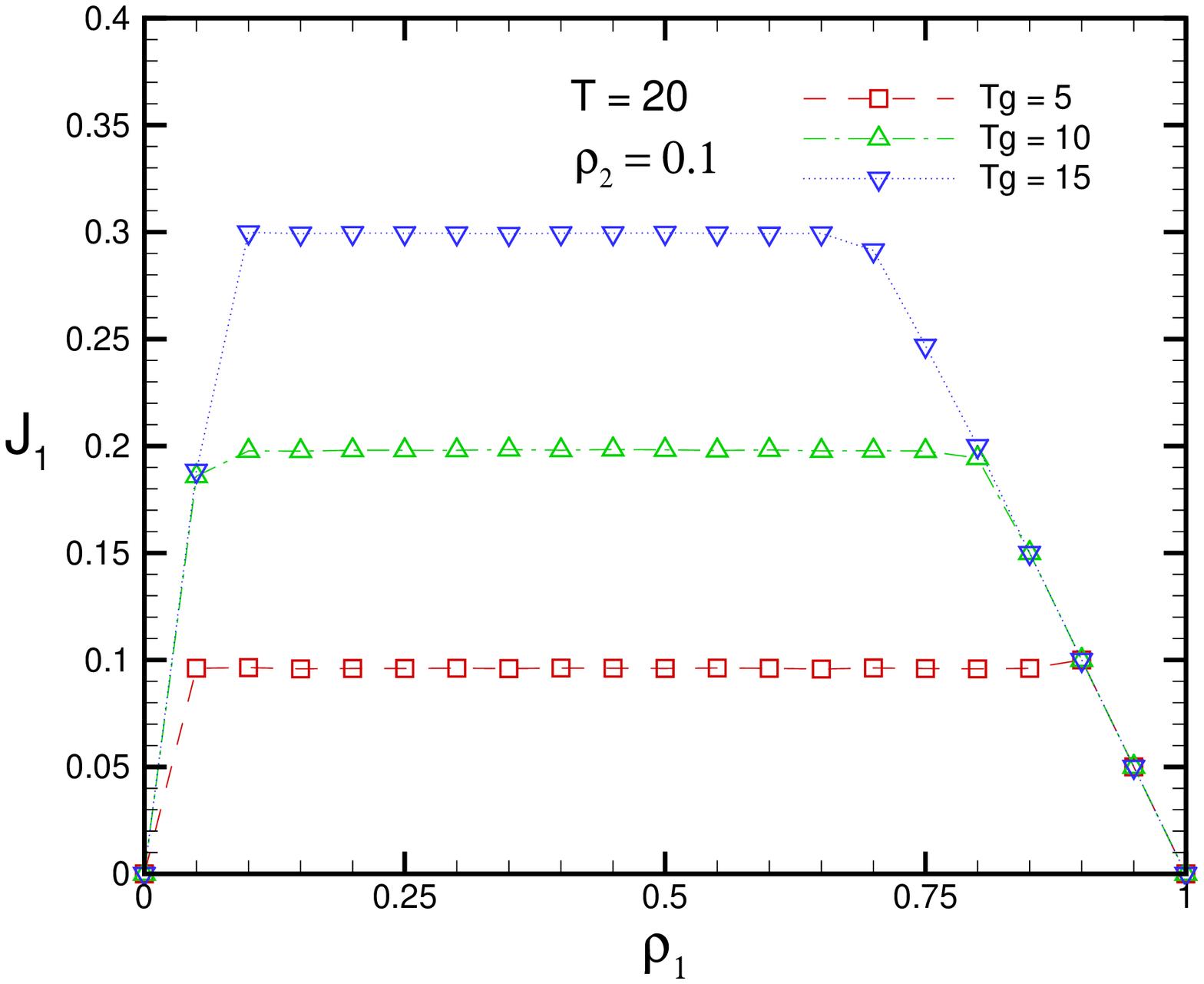}
\includegraphics[width=7cm]{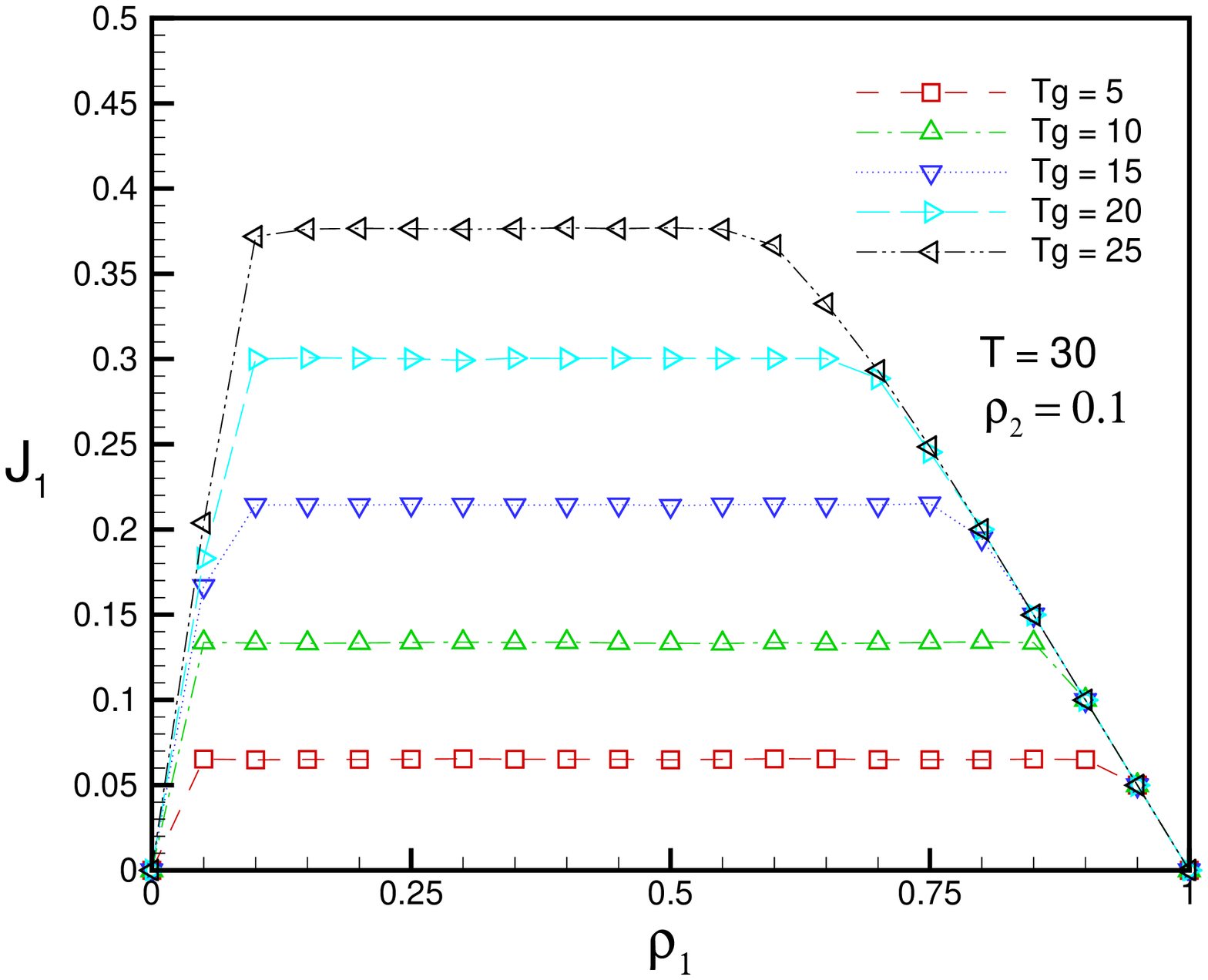}
\includegraphics[width=7cm]{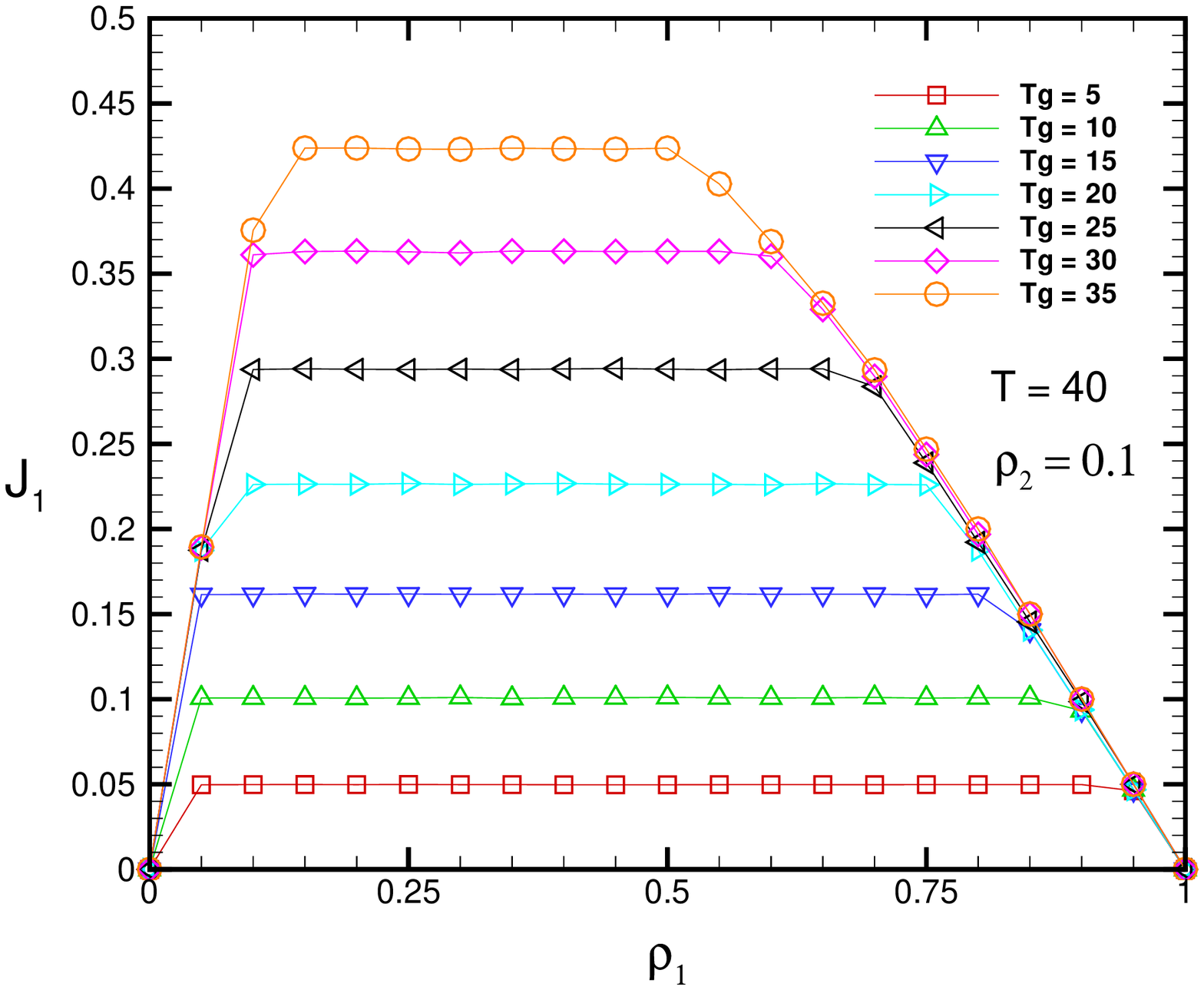}
\caption{ $J_1$ vs $\rho_1$ for various
values of $T_g$ at fixed $\rho_2=0.1$. $T=20$ (top), $T=30$ (middle) and $T=40$ (bottom). } \label{fig:bz2}
\end{figure}

We observe that for each $T_g$, $J_1$ linearly increase up to
$\rho_{1}=\rho_{-}$, then a lengthy plateau region is formed up to
$\rho_{1}=\rho_{+}$. If $T_g$ increases, $\rho_{-}$ becomes
larger, $\rho_{+}$ becomes smaller, and the plateau height
increases. This seems natural because by increasing $T_g$ the
model tends to a normal NS model. The emergence of a plateau
region is associated to defect-like role of the crossing point.
The asymmetric simple exclusion process (ASEP), as a paradigm for
non equilibrium processes in one dimension, with one defective
site has been investigated within the Bethe Ansatz formalism in
\cite{gunter93} and matrix product state \cite{hinrichsen97}. It
is a well-established fact that a local defect can affect the low
dimensional non-equilibrium systems on a global scale. This has
been confirmed not only for ASEP
\cite{lebowitz,barma1,kolomeisky1,foolad6} but also for cellular
automata models describing vehicular traffic flow
\cite{chung,yukawa}. After the plateau, $J_1$ exhibits linear
decrease versus $\rho_1$ in the same manner as in the fundamental
diagram of a single road. Concerning the variation of cycle time
$T$, increasing the cycle time $T$ gives rise, on an equal basis,
to increase both in the green and in the red portion of the cycle
allocated to each street. The results show a notable increase in
flows when $T$ is increased. This observation does not seem to
comply to reality. The reason is due to unrealistic nature of NS
rules. We now consider the flow properties in the second street.
In Figure (3) we depict the behaviour of $J_2$ versus $\rho_1$
for various values of $T_g$.

\begin{figure}
\centering
\includegraphics[width=7cm]{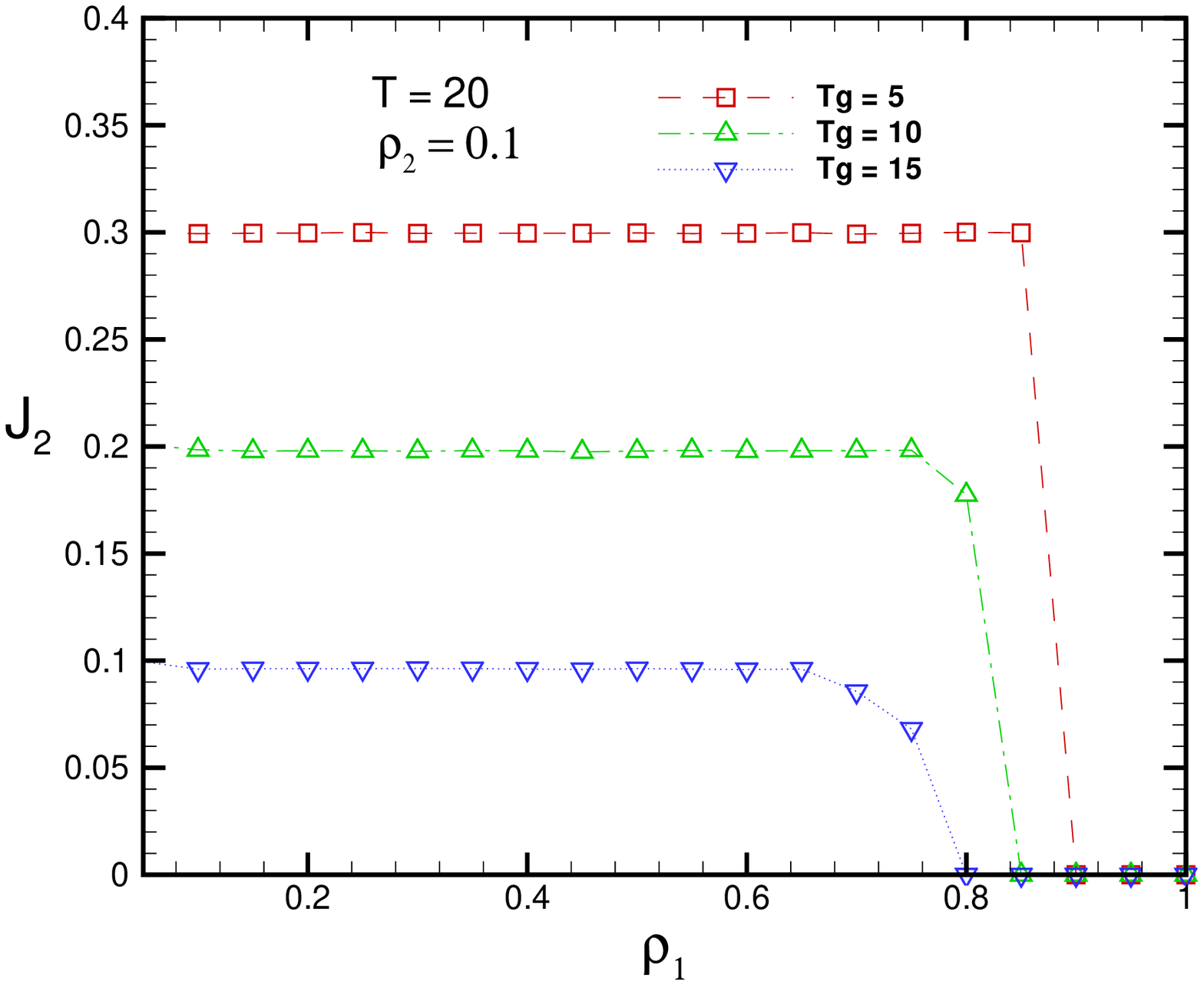}
\includegraphics[width=7cm]{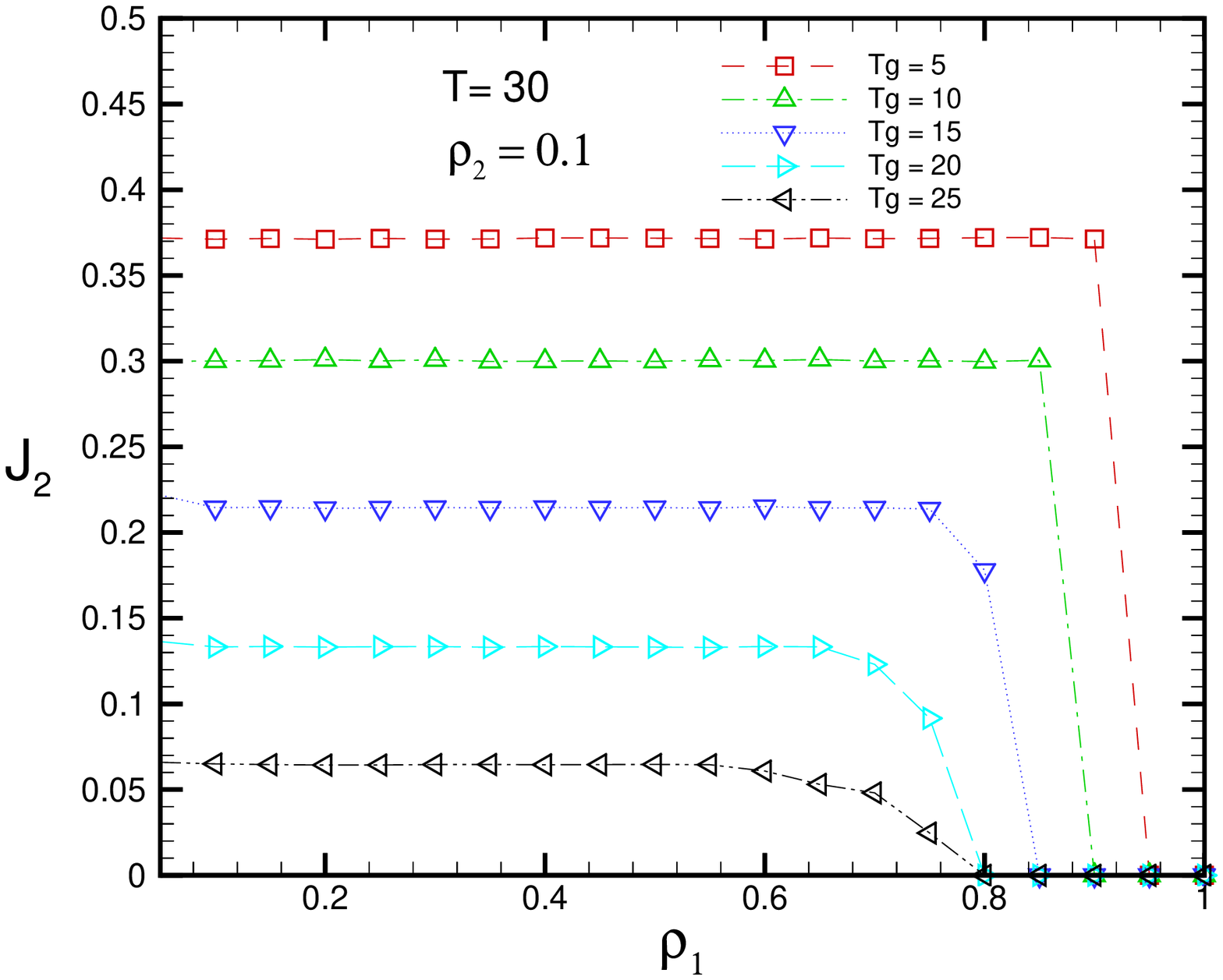}
\caption{ $J_2$ vs $\rho_1$ for various
 $T_g$ at $\rho_2=0.1$. Top: T=20 and bottom: T=30. } \label{fig:bz2}
\end{figure}

Although $\rho_2$ remains constant $J_2$ is affected by density
variations in street $1$. For each $T_g$, $J_2$ as a function of
$\rho_1$ exhibits two regimes. In the first regime, $J_2$ is
almost independent of $\rho_1$ and remains constant up to
$\rho_{+}$. Afterwards in the second regime, $J_2$ exhibits a
linearly decreasing behaviour towards zero. Analogous to $J_1$,
the existence of a wide plateau region indicates that street $2$
can maintain a constant flow capacity for a wide range of density
variations in the first street. If $T_g$ is increased, the green
time allocated to the second street decrease so we expect $J_2$
to exhibit a diminishing behaviour. This is in accordance to what
figure (3) shows. In order to find a deeper insight, it would be
illustrative to look at the behaviour of total current
$J_{tot}=J_1+J_2$ as a function of density in one of the streets.
Evidently for optimisation of traffic one should maximize the
total current $J_{tot}$ therefore it is worth investigating how
the total current behaves upon variation of the model parameters.
Fig.(4) sketches this behaviour.

\begin{figure}
\centering
\includegraphics[width=7cm]{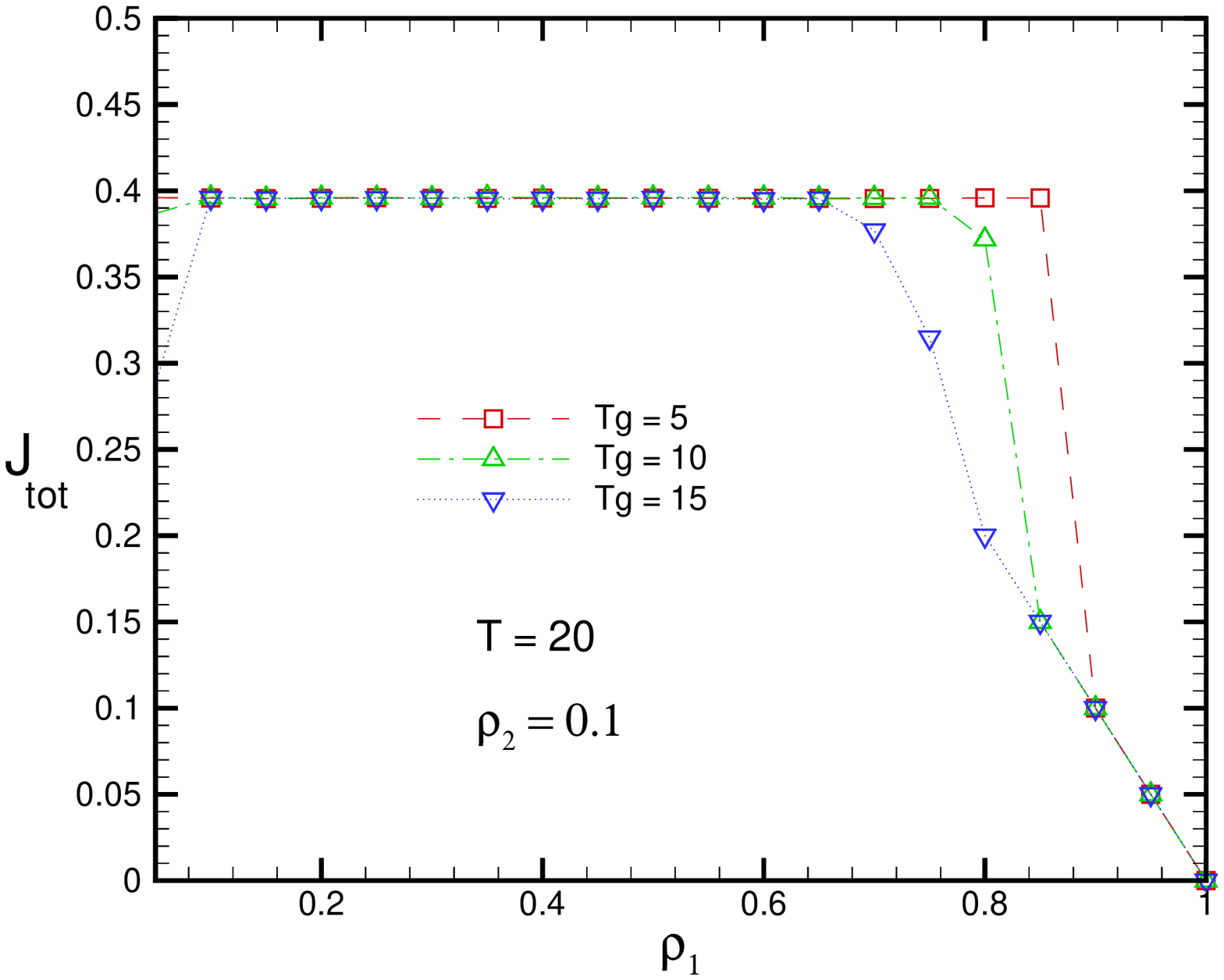}
\includegraphics[width=7cm]{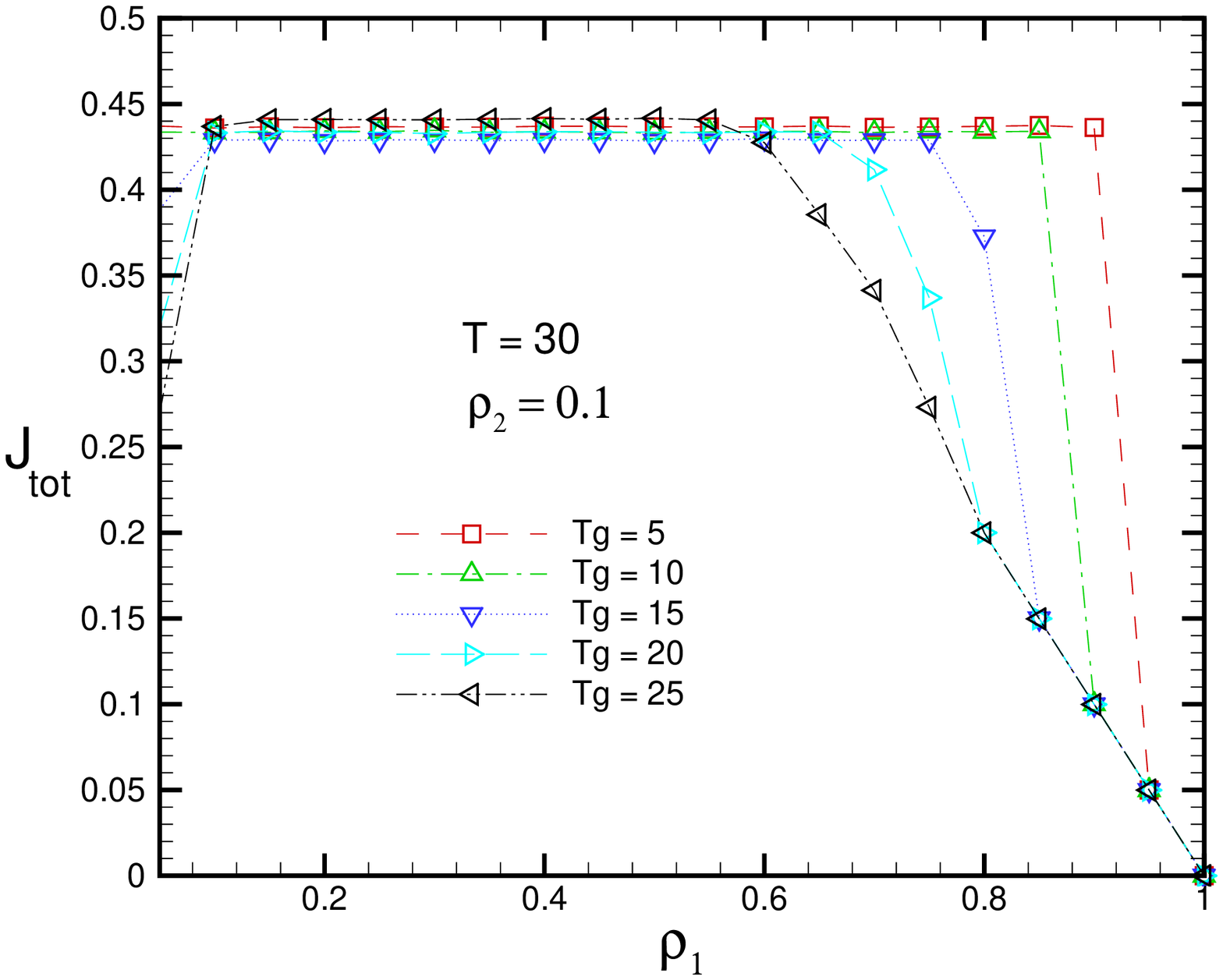}
\includegraphics[width=7cm]{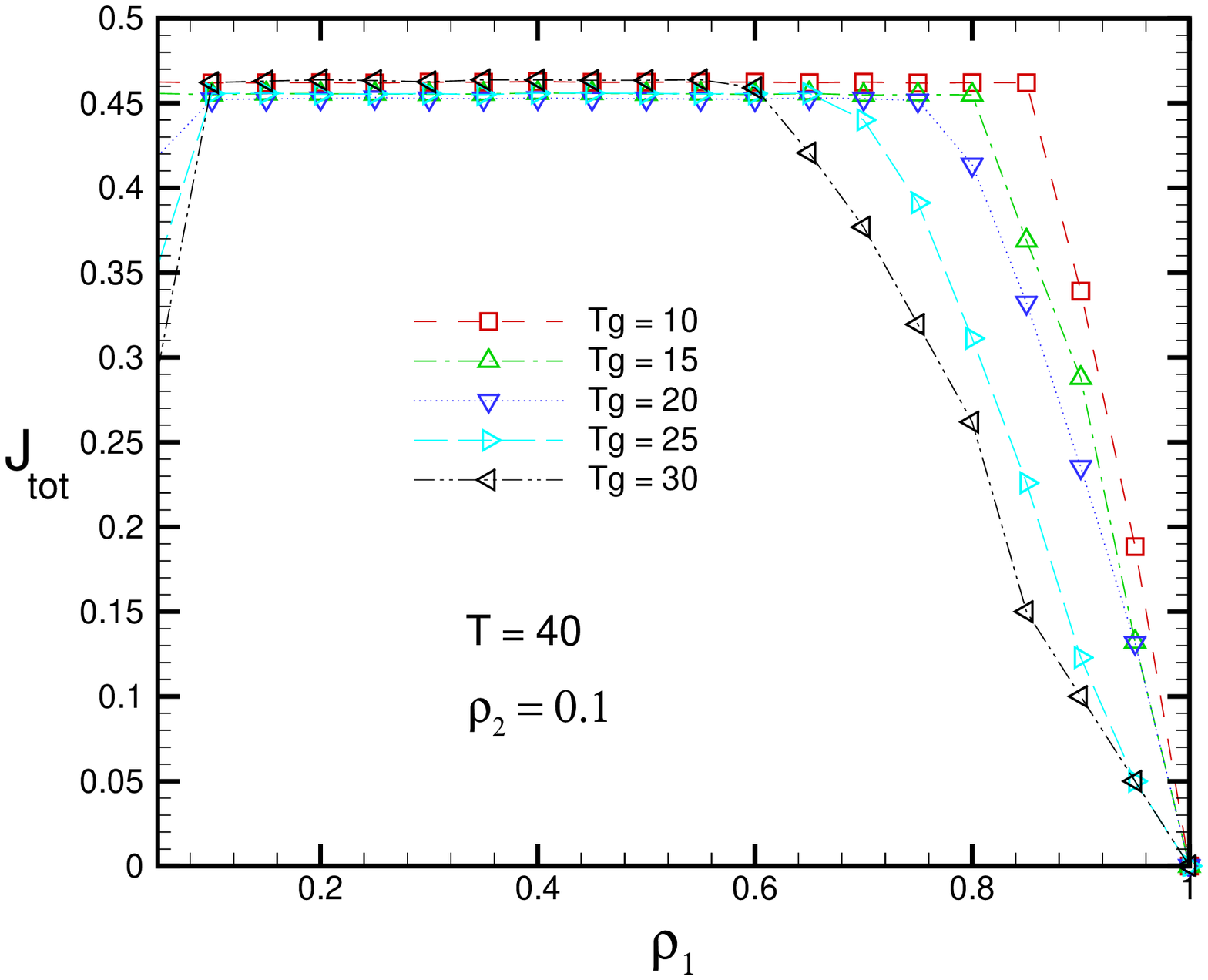}
\caption{ $J_{tot}$ vs $\rho_1$ for various
values of $T_g$ at $\rho_2=0.1$: $T=20$ (Top),
$T=30$ (middle) and $T=40$ (bottom). } \label{fig:bz2}
\end{figure}

In general, the dependence of total current on $J_1$ depends on
the value of $T_g$. Except for small values of $T_g$, total
current increases with $\rho_1$ then it becomes saturated at a
lengthy plateau before it starts its linear decrease. We have also
examined the behaviour of $J_{tot}$ for other values of $\rho_2$.
Figures (5) exhibits the results for $\rho_2=0.05$ and
$\rho_2=0.5$ respectively.

\begin{figure}
\centering
\includegraphics[width=7cm]{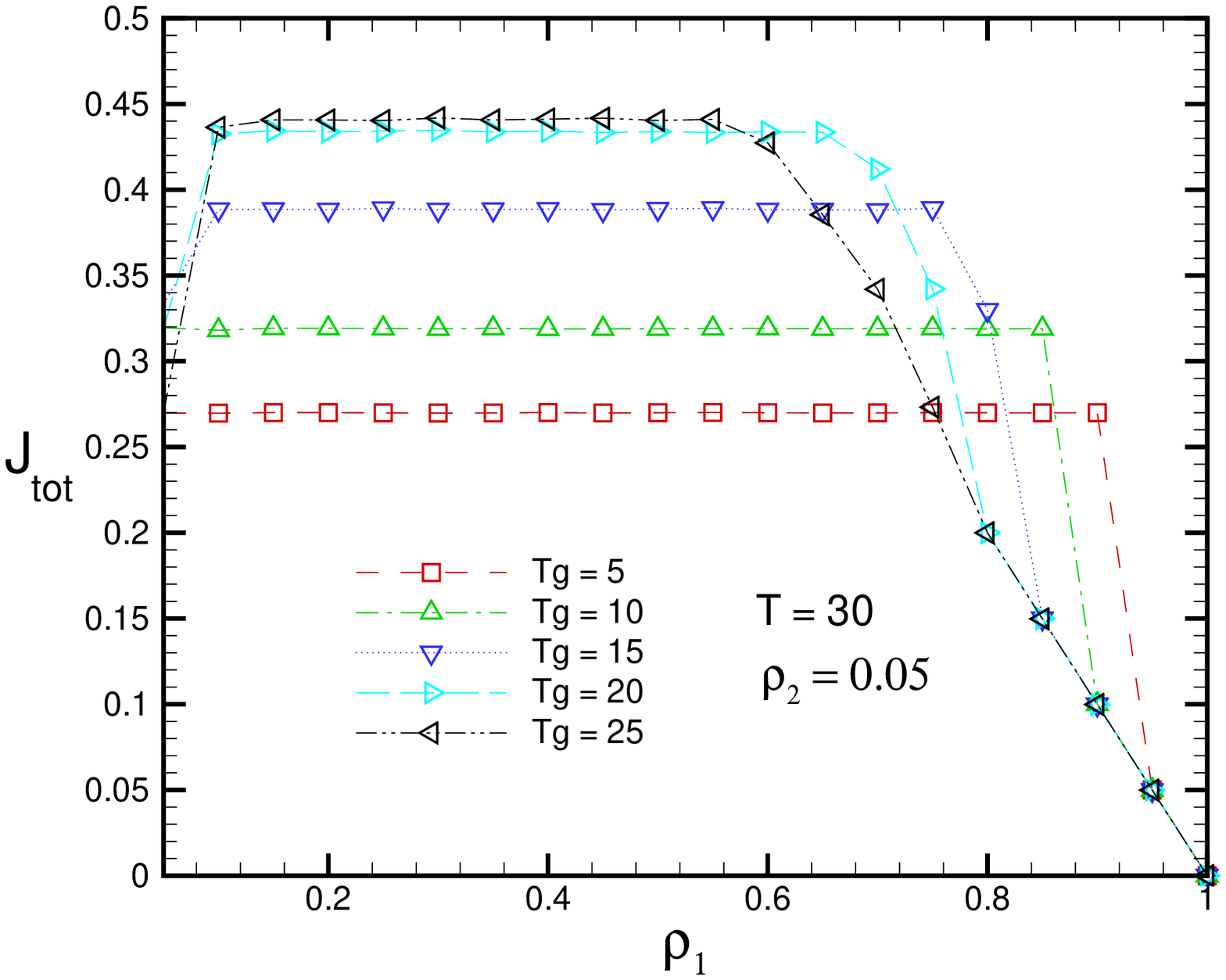}
\includegraphics[width=7cm]{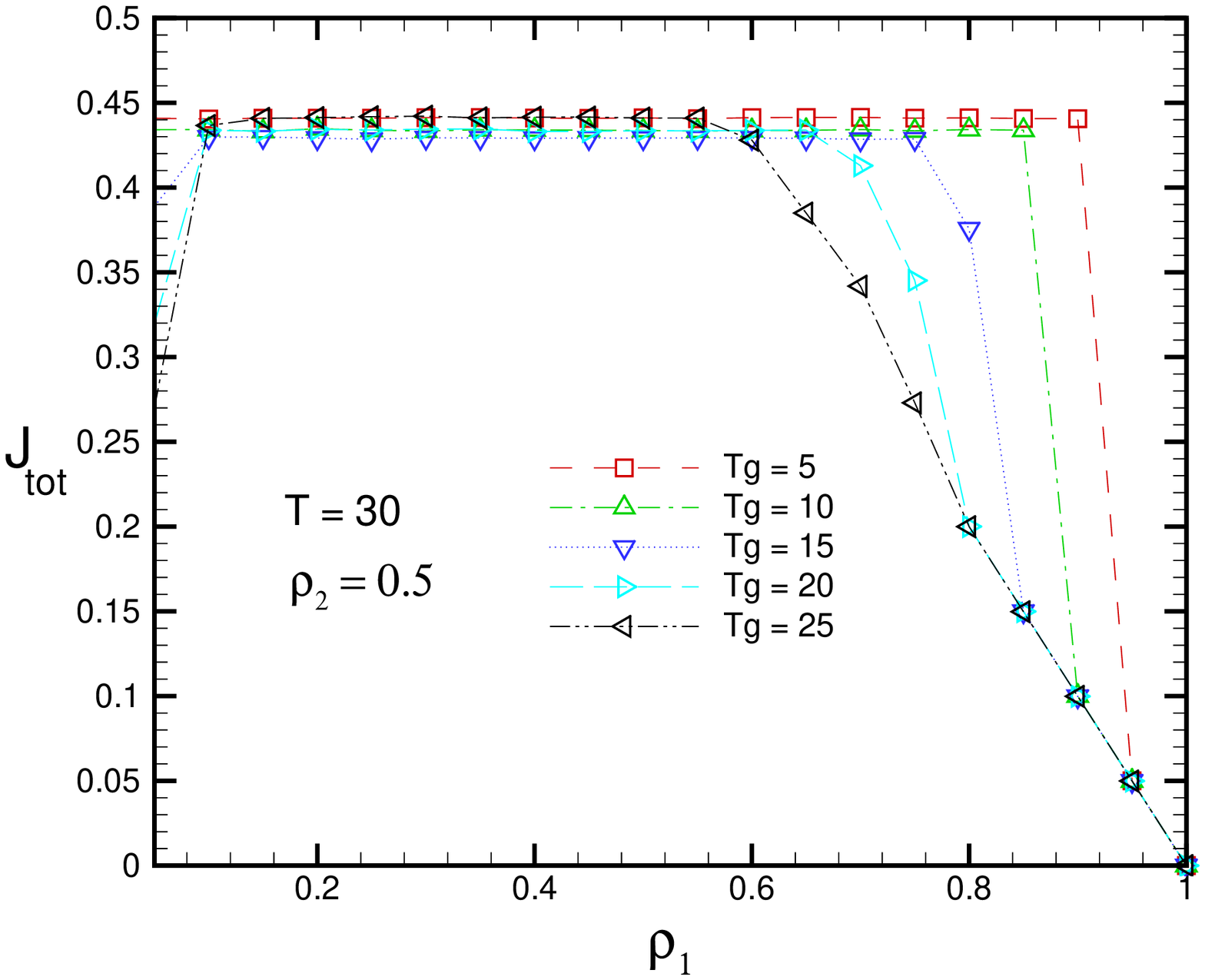}
\caption{ Total current $J_{tot}$ vs
$\rho_1$ for various values of $T_g$ at $T=30$: $\rho=0.05$ (Top) and $\rho=0.5$ (bottom). } \label{fig:bz2}
\end{figure}

Our simulations confirm that for small $\rho_2$ up to $0.1$ total
current shows a distinguishable dependence on $T_g$ in the entire
range of $\rho_1$ especially in intermediate values. In contrast,
for $\rho_2>0.1$, we observe no significant dependence on $T_g$ in
the intermediate $\rho_1$ but we observe notable dependence for
large $\rho_1$.

\subsection{ Density profile and Queue Formation}

In this section we try to investigate, in brief, the
characteristics of queues which are formed behind the crossing
point in the course of the red periods. To this end, in the
following set of figures we exhibit the profile of density in the
vicinity of intersection. This will provide much insight into the
problem. In figure (6) we have sketched such profiles for three
values of green times $T_g$.

\begin{figure}
\centering
\includegraphics[width=7cm]{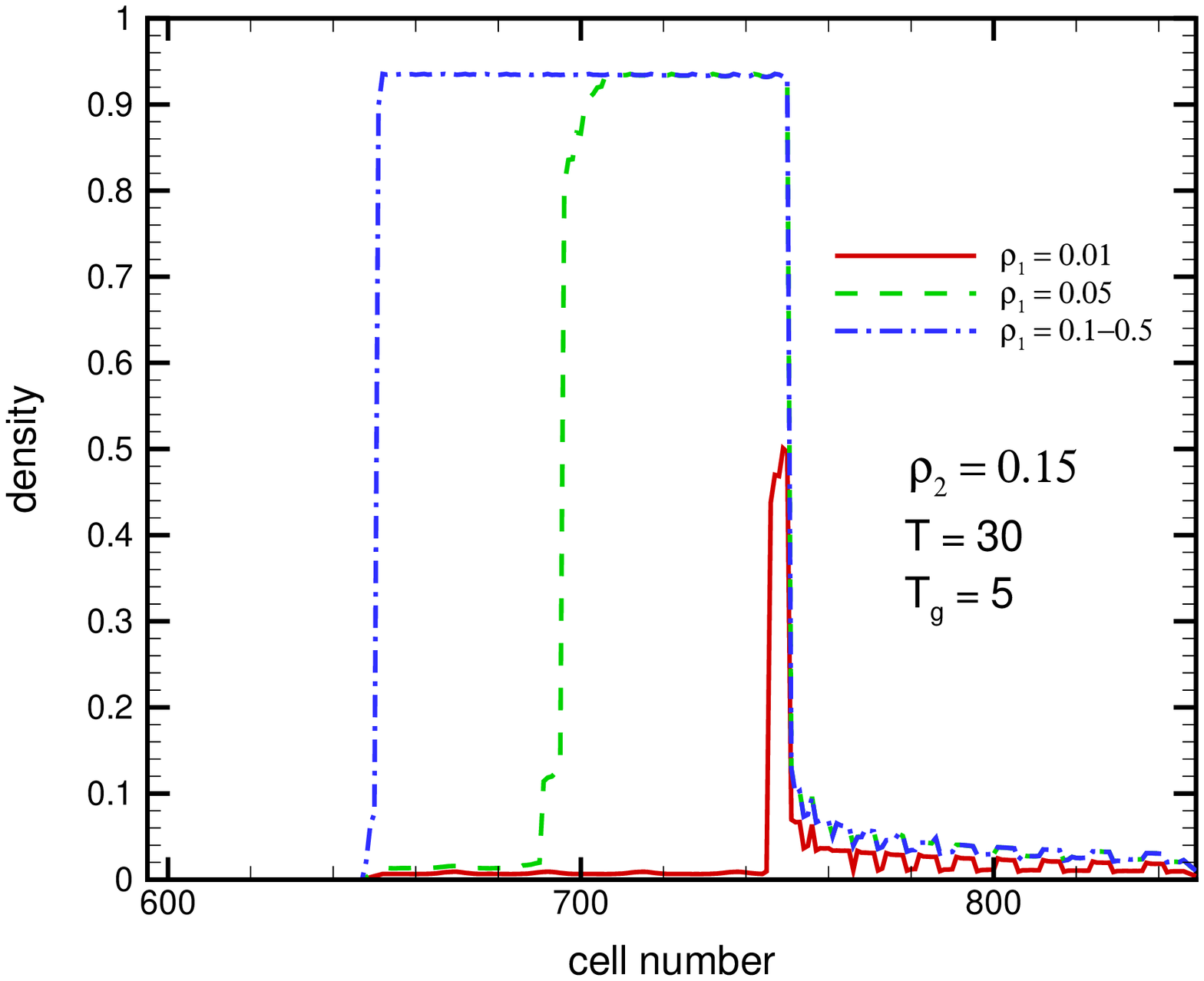}
\includegraphics[width=7cm]{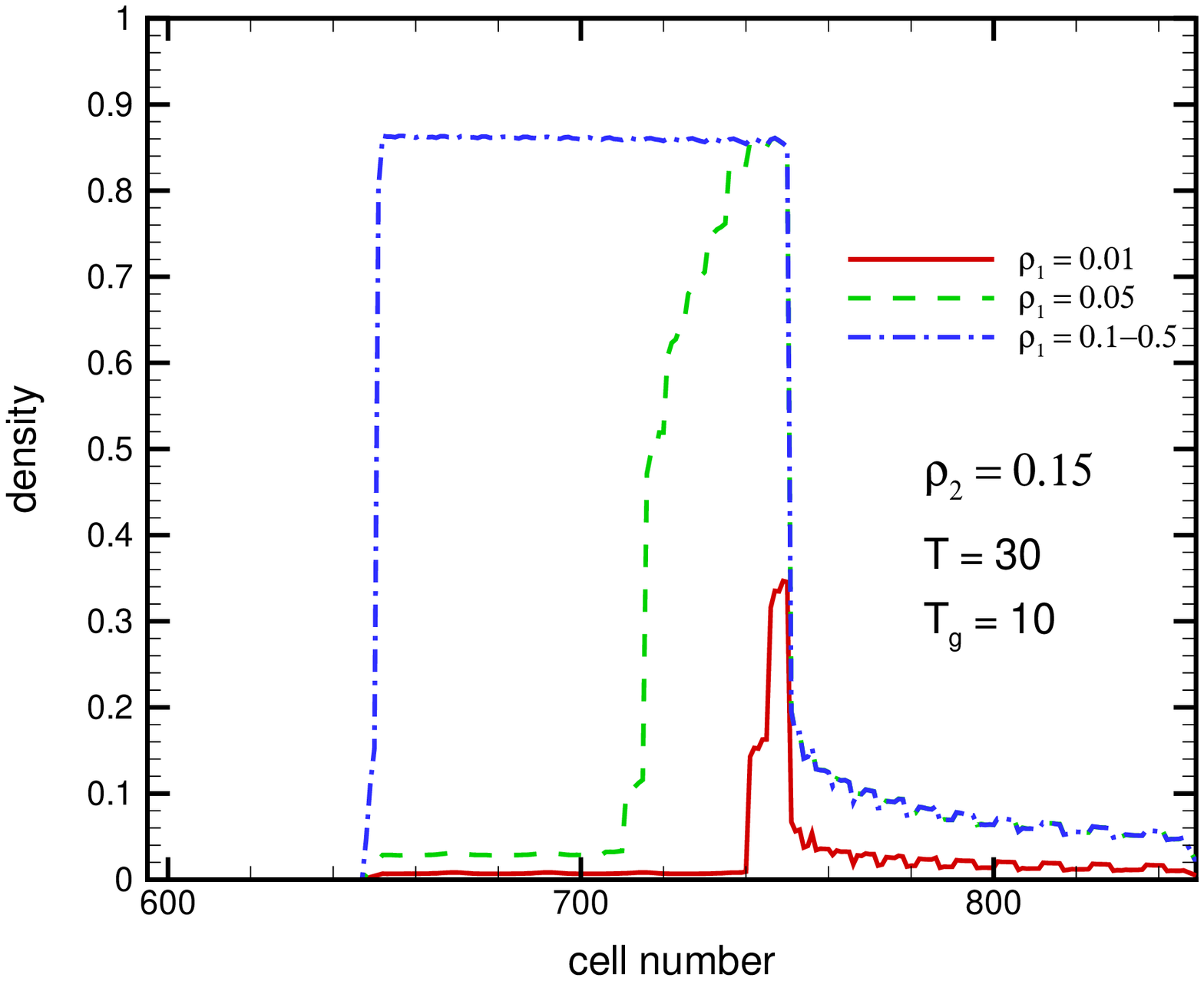}
\includegraphics[width=7cm]{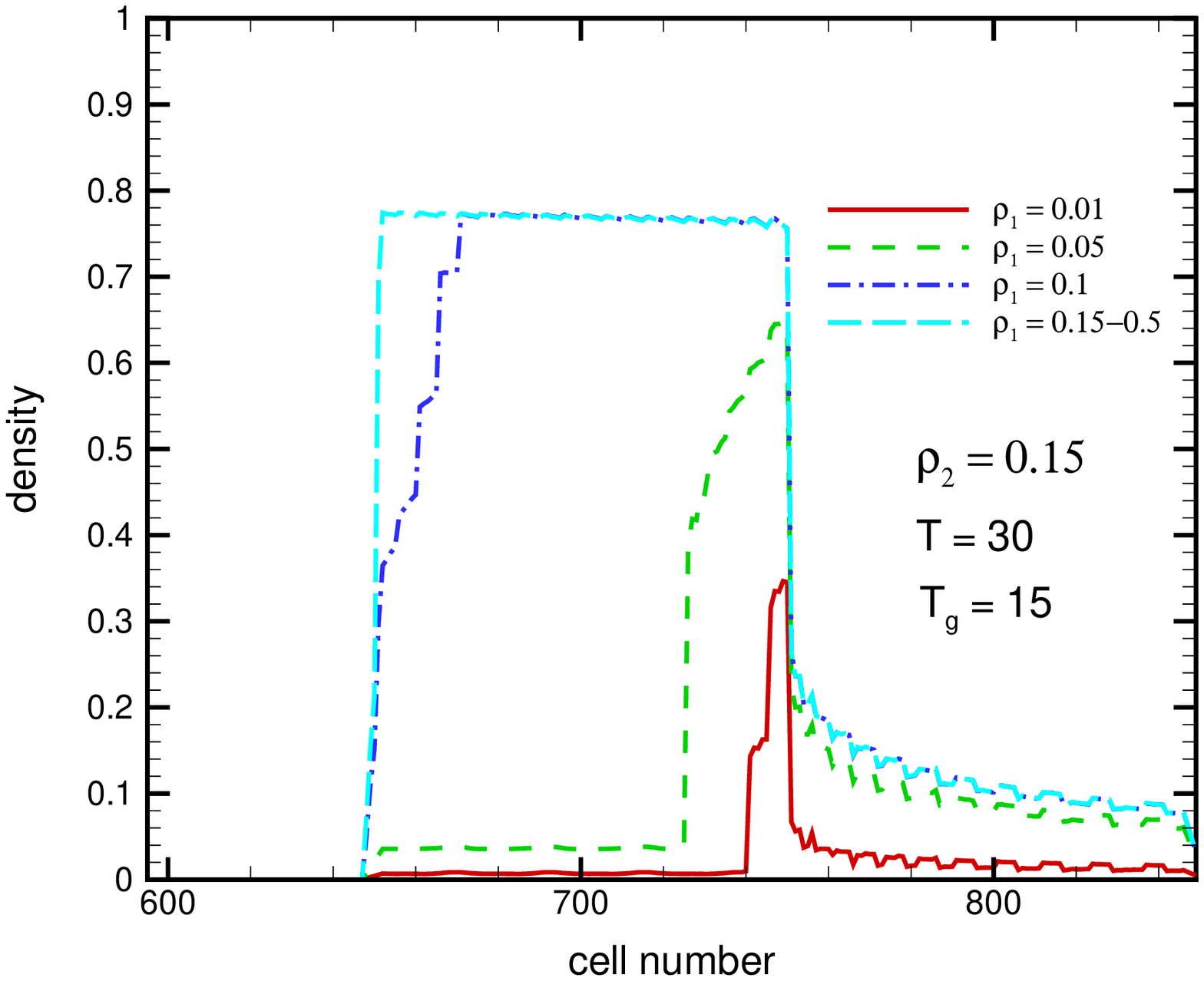}
\caption{ Density profile of the first street for given values of parameters which are
specified in the figures. Crossing site is at $750$th cell. } \label{fig:bz2}
\end{figure}

At each global density $\rho_1$ a high density region, hereafter
referred to as HD, forms behind the crossing point. This
corresponds to queue formation behind the crossing point. The
length and height of this HD region grows with $\rho_1$. Keeping
$ \rho_1$ constant, the length as well as the height of the HD
region decreases with increasing $T_g$. This is expected since by
increasing $T_g$ the cars in the first street are given more time
to pass the intersection and less cars will have to stop at red
periods. Right after the crossing point, we observe a quick
relaxation of the density profile to a low density region
(hereafter referred to as LD). It would be natural to interpret
the length of the HD region as the average queue length. We note
that small oscillation in the profiles are not associated to
inadequacy of simulation time. In fact their emergence is related
to the updating rules of the NS model. Such oscillations have
also been observed in the density-density correlation function in
the NS model \cite{eisenblater}.

\section{ Traffic Responsive Signalisation}

In this section we present our simulations results for the {\it
so-called} intelligent controlling scheme in which the traffic
light cycle is no longer fixed \cite{webster,bell}. In this scheme
which is some times called {\it traffic responsive}, the
signalisation of traffic lights is simultaneously adapted to
traffic status in the vicinity of intersection. This scheme has
been implemented in simulation of traffic flow at intersections
with open-boundary conditions \cite{foolad2,foolad4}. There exist
numerous schemes in which traffic responsive signalisation can be
prescribed. Three of these schemes are discussed in
\cite{foolad2}. Here for brevity we discuss only one of these
methods. To be precise, we define a cut-off queue length $Q_c$.
The signal remain red for a street until the length of the
corresponding queue formed behind the red light exceeds the
cut-off length $Q_c$. At this moment the lights change colour.
Apparently due to stochastic nature of cars movement, the cycle
time will be subjected to variations and will no longer remain
constant. In figure (7) we exhibit $J_{tot}$ versus $\rho_1$ for
various values of cut-off lengths $Q_c$ and $\rho_2$.

\begin{figure}
\centering
\includegraphics[width=7cm]{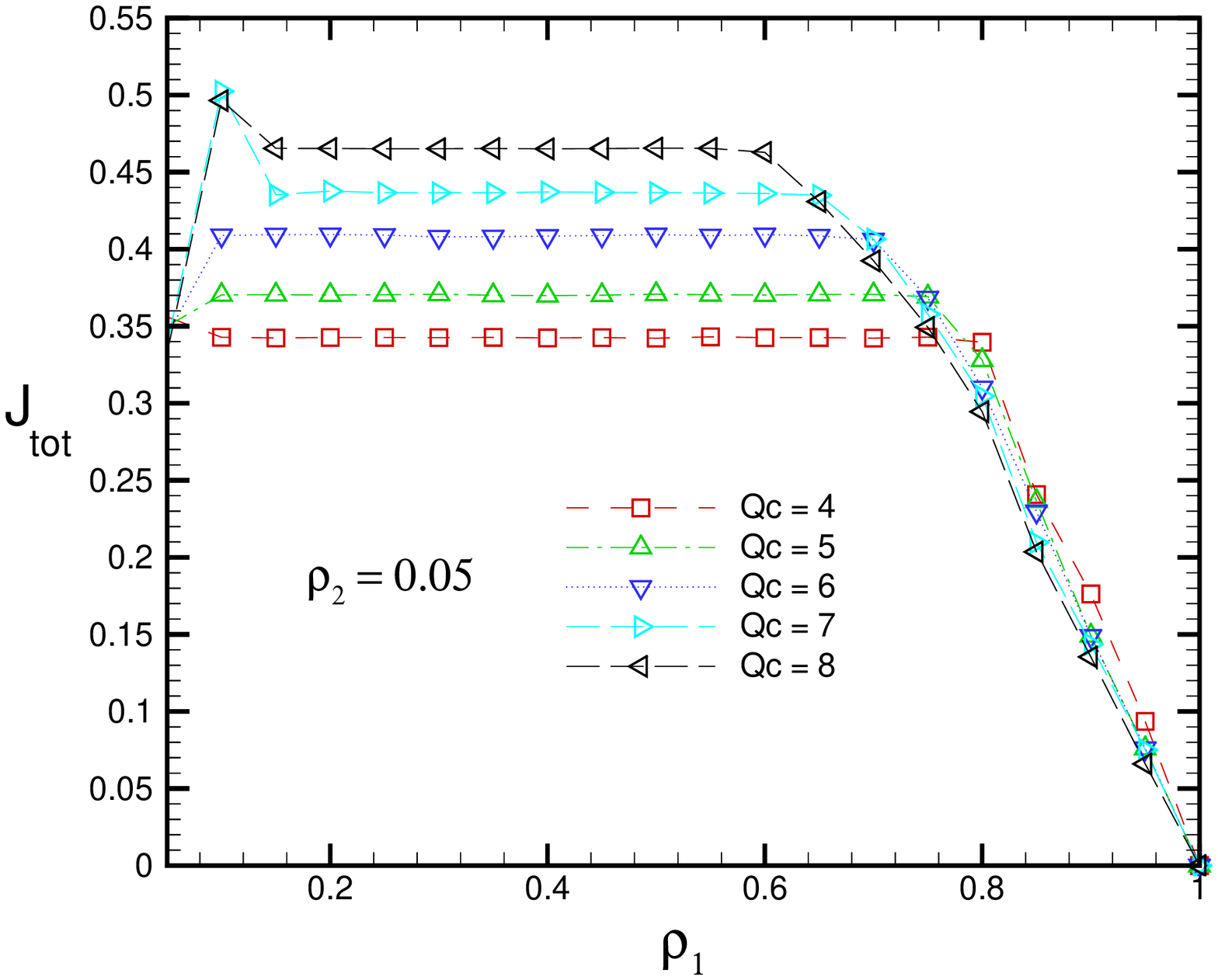}
\includegraphics[width=7cm]{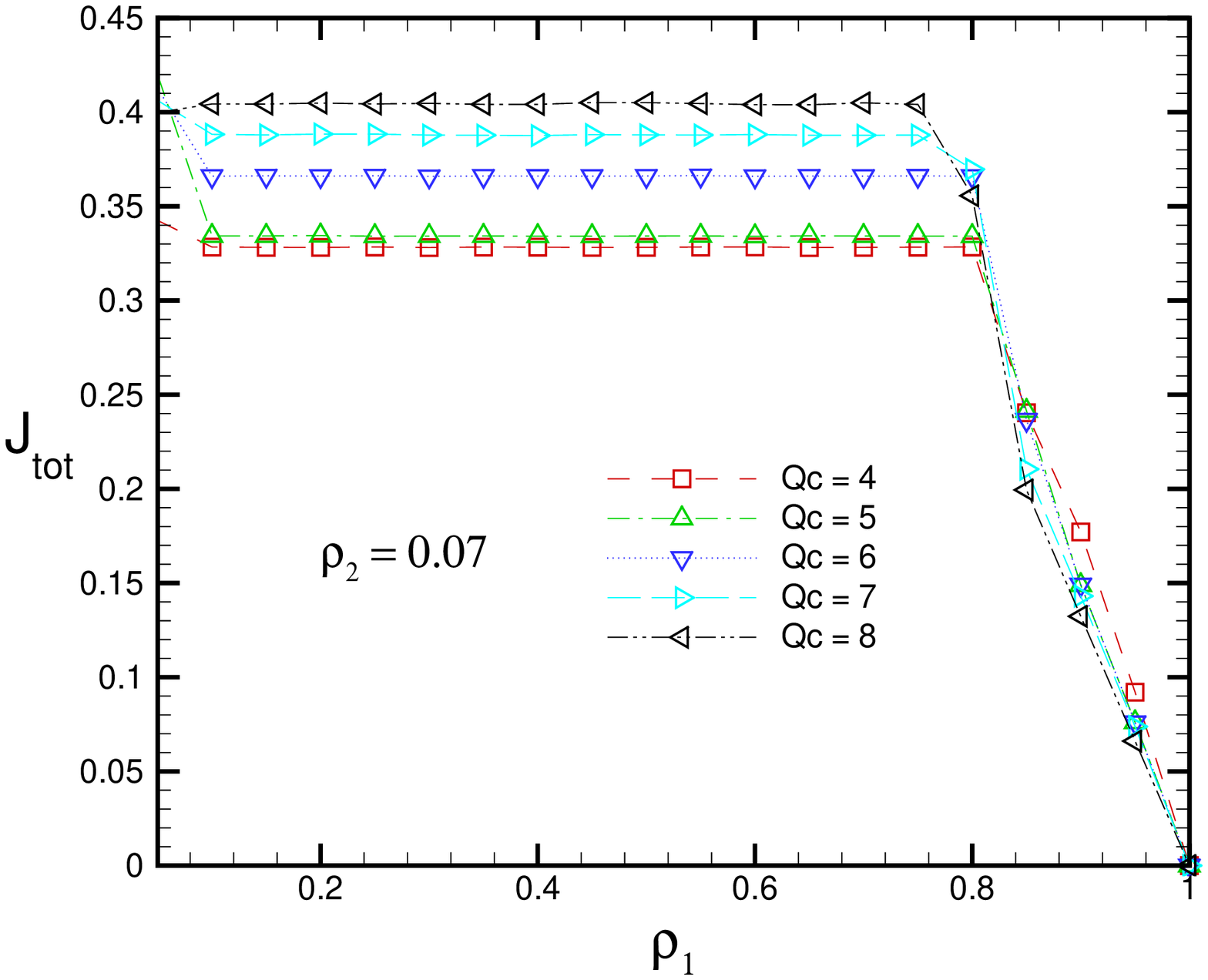}
\caption{ Total current vs $\rho_1$ at $\rho_2=0.05$ (top) and $\rho_2=0.07$ (bottom)
for various values of $Q_c$. } \label{fig:bz2}
\end{figure}

Analogous to fixed-time scheme, for given $\rho_2$ a lengthy
plateau in $J_{tot}$ forms. The plateau height as well as its
length show a significant dependence on $Q_c$. higher $Q_c$ are
associated with smaller length and higher current. We have also
examined larger values of $\rho_2$. The results are qualitatively
analogous the above graphs. The notable point is that for $\rho_2$
larger than 0.1, $J_{tot}$ do not show a significant dependence
on $\rho_2$. Figure (8) depicts $J_{tot}$ versus $\rho_1$ for
$\rho_2=0.15$.

\begin{figure}
\centering
\includegraphics[width=7cm]{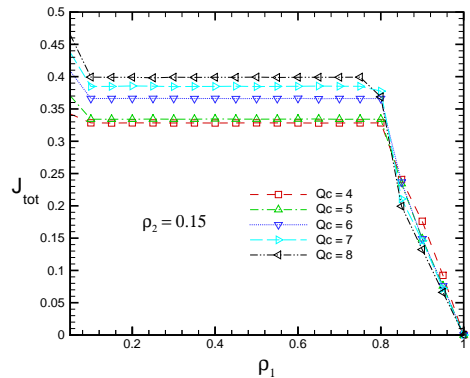}
\caption{ $J_{tot}$ vs $\rho_1$ at $\rho_2=0.15$ for various values of $Q_c$. } \label{fig:bz2}
\end{figure}

In figure (9) we have depicted the cycle lengths in the traffic
responsive scheme which vary from cycle to cycle. The appearance
of such behaviour remarks the adaptation of traffic lights in the
responsive scheme.

\begin{figure}
\centering
\includegraphics[width=7cm]{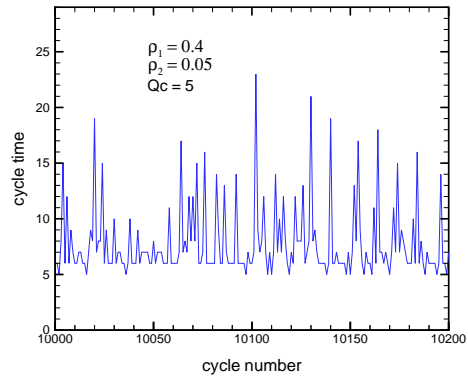}
\caption{ Variation of cycle lengths in traffic responsive scheme for 200 cycles. } \label{fig:bz2}
\end{figure}

To shed some light onto the problem, we sketch space-time plots
of vehicles. It is seen that in traffic responsive scheme, the
cars spatial distribution is more homogeneous which is due to
randomness in cycle times.

\begin{figure}
\centering
\includegraphics[width=7cm]{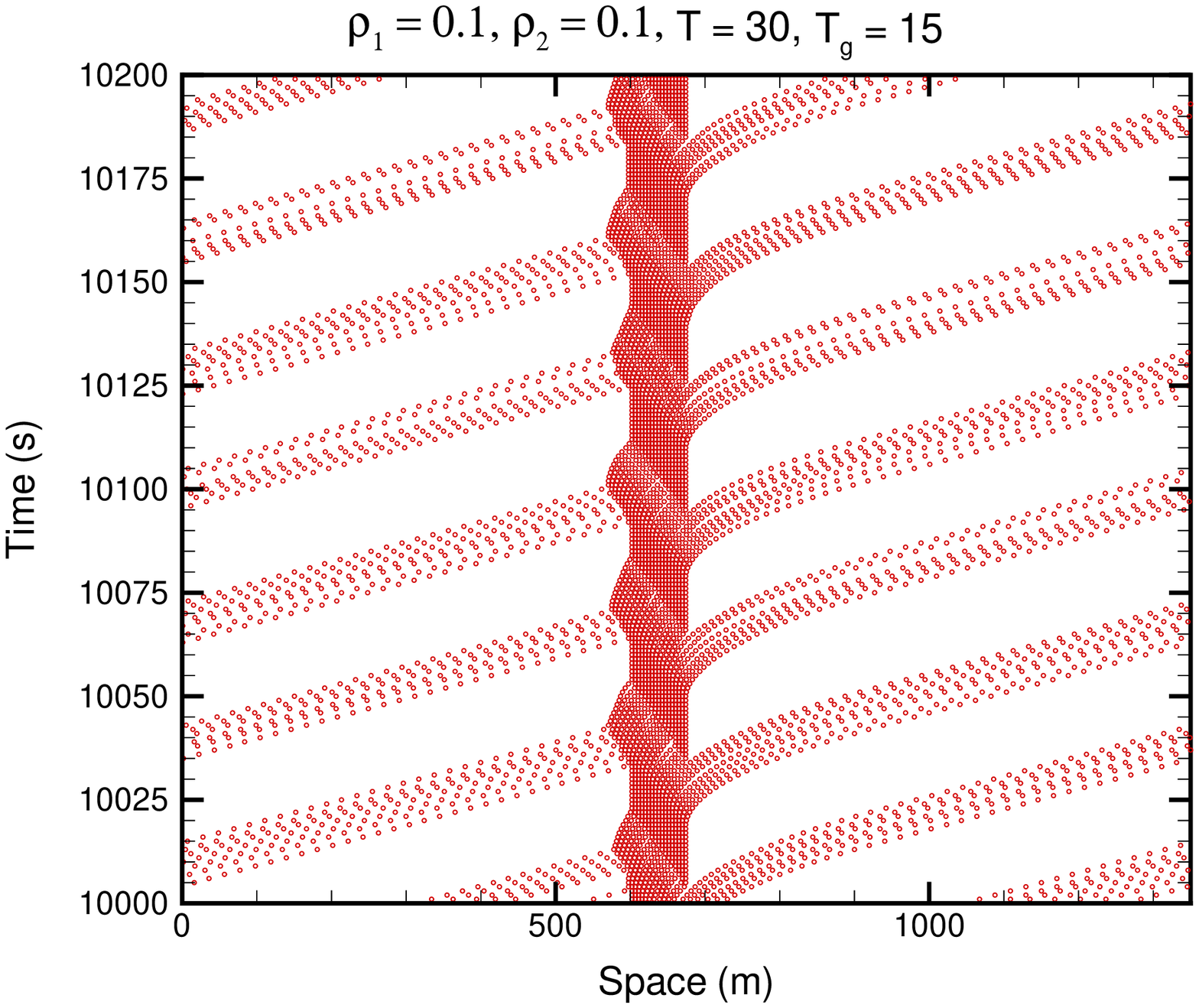}
\includegraphics[width=7cm]{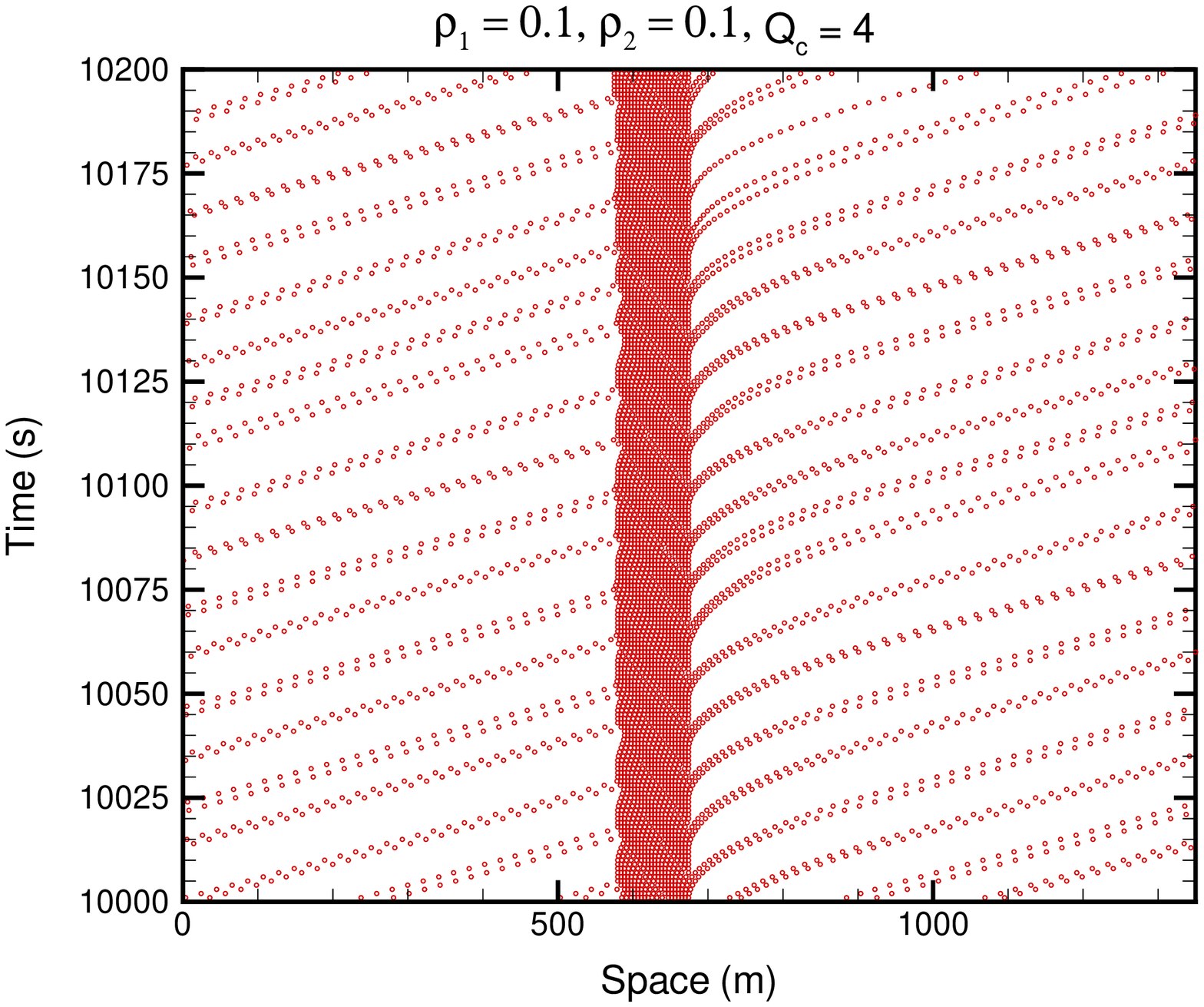}
\caption{ Space-time plot of vehicles for fixed time (top) and traffic responsive schemes (bottom). } \label{fig:bz2}
\end{figure}

Lastly, we compare our results to those obtained in simulation of
a nonsignalised intersection \cite{foolad5}. In a nonsignalised
intersection, the cars yield to each other upon approaching the
crossing point and the priority is give to the car which is closer
to the crossing point. The total current versus $\rho_{1}$ in a
nonsignalised intersection has the following form:

\begin{figure}
\centering
\includegraphics[width=7cm]{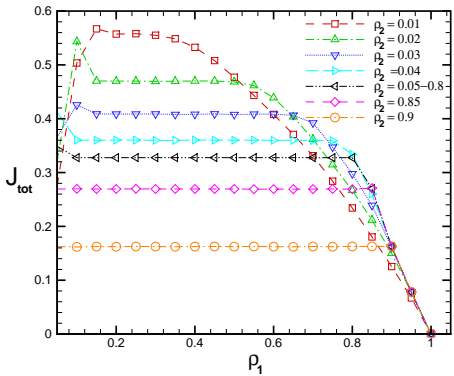}
\caption{ $J_{tot}$ vs $\rho_1$ for various values of $\rho_2$ in a nonsignalised intersection. } \label{fig:bz2}
\end{figure}

For $\rho_2$ larger than $0.1$, the optimal current is roughly
$0.33$ in the nonsignalisation scheme whereas in the signalised
scheme with a fixed time strategy, the optimal value of $J_{tot}$
reaches beyond 0.45 (see Figs. 4 and 5) in a fixed time scheme and
0.4 in the traffic responsive scheme. It can be concluded that
signalisation strategies are apparently more efficient in comparison
to non-signalisation scheme.\\

\section{Summary and Concluding Remarks}

By extensive Monte Carlo simulations, we have investigated the flow
characteristics in a signalised intersection via developing a
Nagel-Schreckenberg cellular automata model. We have considered two
types of schemes: fixed-time and traffic responsive. In particular,
we have obtained the fundamental diagrams in both streets and the
dependence of total current on street densities. Our findings show
hindrance of cars upon reaching the red light gives rise to
formation of plateau regions in the fundamental diagrams. This is
reminiscent of the conventional role of a single impurity in the one
dimensional out of equilibrium systems. The existence of wide
plateau region in the total system current shows the robustness of
the controlling scheme to the density fluctuations. The overall
throughput from the intersection shows a significant dependence on
the cycle time in the fixed time scheme and on the queue cut-off
length in the responsive scheme. Moreover, by plotting the density
profiles, we have been able to quantify the characteristics of
queues. Comparison to our previous results for a nonsignalised
intersection shows the higher efficiency of the signalisation
strategy. We remark that our approach is open to serious challenges.
The lack of empirical data prevents us to judge how much our results
are close to and to what degree they differ from reality. Our CA
model allows for varying space and time grids. By their appropriate
adjusting, we are able to reproduce a realistic acceleration. We
emphasize that our model suffers from non realistic behaviour of car
movement when the lights turn green. The crucial point is to model
the braking and accelerating as realistic as possible. Empirical
data are certainly required for this purpose. We expect the system
characteristics undergo substantial changes if realistic yielding
declaration is taken into account.

\section{acknowledgement}

We highly appreciate M. Abol Bashar for his fruitful helps and
enlightening discussions.


\begin{thebibliography}{99}


\bibitem{schadrev} D. Chowdhury, L. Santen and A. Schadschneider, {\em Physics Reports}, {\bf 329}, 199 (2000).



\bibitem{helbingrev} D. Helbing, {\em Rev. Mod. Phys.}, {\bf 73}, 1067 (2001).



\bibitem{kernerbook} {\em Physics of traffic flow}, B. Kerner, Springer (2004).






\bibitem{tgf05} \  R. K\"{u}ne, A. Schadschneider,  M.\ Schreckenberg and D.E.\ Wolf (eds.) {\em Traffic and Granular
flow 05}, (Springer, 2007).



\bibitem{bml} O.\ Biham, A.\ Middleton and D.\ Levine, {\em Phys. Rev. A}, {\bf 46}, R6124 (1992).



\bibitem{nagatani1} T. Nagatani, {\em J. Phys. Soc. Japan}, {\bf 63}, 1228
(1994); {\em J. Phys. Soc. Japan}, {\bf 64}, 1421 (1995).



\bibitem{nagatani2} T. Nagatani, {\em Phys. Rev. E}, {\bf 51}, 922
(1995); and T. Seno, {\em Physica A}, {\bf 207}, 574 (1994).



\bibitem{cuesta} J.A. Cuesta, F.C. Martines, J.M. Molera and A. Sanchez,
{\em Phys. Rev. E}, {\bf 48}, R4175 (1993).




\bibitem{freund} J. Freund and T. P\"oschel, {\em Physica A}, {\bf 219}, 95 (1995).



\bibitem{chopard} B. Chopard, P.O. Luthi and P.A. Queloz {\em J. Phys. A}, {\bf 29}, 2325 (1996).




\bibitem{tadaki} S. Tadaki, {\em Phys. Rev. E}, {\bf 54}, 2409
(1996); J. Physi. Soc. Jpn. {\bf 66}, 514 (1997).




\bibitem{torok} J. T\"or\"ok and J. Kertesz, {\em Physica A}, {\bf 231},
515 (1996).



\bibitem{chau} H.F. Chau, K.Y. Wan and K.K. Yan, {\em Physica A}, {\bf 254},
117 (1998).



\bibitem{cs} D.\ Chowdhury and A.\ Schadschneider, {\em Phys. Rev. E}, {\bf 59}, R 1311 (1999).



\bibitem{brockfeld} E.\ Brockfeld, R.\ Barlovic, A.\ Schadschneider and and M.\
Schreckenberg, {\em Phys. Rev. E}, {\bf 64}, 056132 (2001).



\bibitem{chitur} Y. Chitur and B. Piccoli, {\em Discrete and
Continuous Dynamical Systems B}, {\bf 5}, 599 (2005).



\bibitem{nagatani3} T. Nagatani, {\em J. Phys. A}, {\bf 26}, 6625 (1993).



\bibitem{ishibashi1} Y. Ishibash and M. Fukui, {\em J. Phys. Soc. Japan}, {\bf 65},
2793 (1996).



\bibitem{ishibashi2} Y. Ishibash and M. Fukui, {\em J. Phys. Soc. Japan}, {\bf 70}, 2793 (2001);
{\em J. Phys. Soc. Japan}, {\bf 70}, 3747 (2001).


\bibitem{foolad1} M.E. Fouladvand and M. Nematollahi, {\em Eur. Phys. J. B}, {\bf 22}, 395 (2001).



\bibitem{krbalek} M. Krbalek and P. Sebra, {\em J. Phys. A }, {\bf 36},  L7 (2003).



\bibitem{foolad2} M. E. Fouladvand, Z. Sadjadi and M. R. Shaebani {\em J. Phys. A: Math. Gen}, {\bf 37}, 561 (2004).



\bibitem{foolad3} M. E. Fouladvand, Z. Sadjadi and M. R. Shaebani {\em Phys. Rev. E}, {\bf 70}, 046132  (2004).



\bibitem{foolad4} M. E. Fouladvand, M. R. Shaebani and Z. Sadjadi {\em J. Phys. Soc. Japan}, {\bf 73}, No. 11, 3209 (2004).



\bibitem{helbing1} D. Helbing, S. L$\ddot{a}$mmer and J.P. Lebacque, in C. Deissenberg and R.F. Hartl(eds.),
{\em Optimal Control and Dynamic Games}, p. 239, Springer,
Dortrecht, 2005; arXive physics/0511018.



\bibitem{helbing2} S. L$\ddot{a}$mmer, H. Kori, K. Peters and D.
Helbing, {\em Physica A}, {\bf 363}, 39 (2006).



\bibitem{helbing3} R. Jiang, D. Helbing, P. Kumar Shukla and Q-S Wu;
{\em Physica A}, {\bf 368}, issue~2, 567 (2006).



\bibitem{ray} B. Ray and S.N. Bhattacharyya, {\em Phys. Rev. E}, {\bf 73}, 036101 (2006).



\bibitem{chen} R. X. Chen, K. Z. Bai and M. R. Liu, {\em Chinese Physics}, {\bf 15}, Issue~7, 1471 (2006).



\bibitem{wang} R. Wang, M. Liu, R. Kemp and M. Zhou, {\em Int. J. Mod. Phys. C}, {\bf 18}, issue~5, 903 (2007).



\bibitem{huang} D. W. Huang, {\em Physica A}, {\bf 383}, Issue 2, 603 (2007).



\bibitem{najem} M. Najem, {\em Int. J. Mod. Phys. C}, {\bf 18}, Issue~6, 1047 (2007).



\bibitem{foolad5} M. E. Foulaadvand and S. Belbasi, {\em J. Phys. A: Math, Theor.}, {\bf 40}, 8289 (2007).



\bibitem{foolad6} M. E. Foulaadvand and M. Neek Amal, {\em Euro. Phys. Lett.}, {\bf 80}, issue 6, Article number 6002 (2007).



\bibitem{ns} K. Nagel, M. Schreckenberg, {\em J.Phys. I France}, {\bf 2}, 2221 (1992).



\bibitem{gunter93} G. M. Sch\"{u}tz, {\em J. Stat. Phys.}, {\bf 71}, 471 (1993).



\bibitem{hinrichsen97} H. hinrichsen and S. Sandow {\em J. Phys. A: Math. Gen}, {\bf 30}, 2745 (1997).



\bibitem{lebowitz} S. Janowsky and J. Lebowitz {\em Phys. Rev. A}, {\bf 45}, 618 (1992).



\bibitem{barma1} G. Tripathy and M. Barma, {\em Phys. Rev. Lett.}, {\bf 78}, 3039 (1997).



\bibitem{kolomeisky1} A. B. Kolomeisky, {\em J. Phys. A: Math, Gen.}, {\bf 31}, 1153 (1998).



\bibitem{chung} K.H. Chung and P.M. Hui, {\em J. Phys. Soc. Jap.}, {\bf 63}, 4338 (1994).



\bibitem{yukawa} S. Yukawa, M. Kikuchi and S. Tadaki {\em J. Phys. Soc. Jap.}, {\bf 63}, 3609 (1994).



\bibitem{eisenblater} B. Eisenbl\"{a}tter, L. Santen, A.
Schadschneider and M. Schreckenberg, {\em Phys. Rev. E}, {\bf
57}, 1309 (1998).



\bibitem{webster} F. Webster and B. Cobb in {\em Traffic Signal
Setting}, (H.M.S Office, London, 1966).



\bibitem{bell} M. G. Bell, {\em Transp. Res. A}, {\bf 26}, No.4,
303 (1992).


\end{thebibliography}
\end{document}